\documentclass{iau}
\usepackage{graphicx}

\title[AGN disk-jet luminosity correlation] 
{On the relation between the AGN jet and accretion disk emissions}

\author[Vahe' Petrosian \& Jack Singal]   
{
Vahe' Petrosian$^1$
 \and 
Jack Singal$^2$
}

\affiliation{
$^1$Dept. of Physics and KIPAC, Stanford University, \\ 
382 Via Pueblo Mall, Stanford, CA 94306, USA \\ 
email: {\tt vahep@stanford.edu} \\
[\affilskip]
$^2$Dept. of  Physics,  University of Richmond, \\ 
Richmond, VA 23173\\
email: {\tt jsingal@richmond.edu}
}

\pubyear{2014}
\volume{313} 
\pagerange{xx--xx}
\jname{Extragalactic jets from every angle}
\editors{F. Massaro, C.C. Cheung, E. Lopez, A. Siemiginowska, eds.}

\def\beq{\begin{equation}}
\def\eeq{\end{equation}}

\begin{document}

\maketitle

\begin{abstract}

\end{abstract}
Active galactic nuclei jets are detected via their radio and/or gamma-ray emissions while the accretion disks are detected by their optical and UV radiation. Observations of the radio and optical luminosities show a strong correlation between the two luminosities. However, part of this correlation is due to the redshift or distances of the sources that enter in calculating the luminosities from the observed fluxes and part of it could be due to the differences in the cosmological evolution of luminosities. Thus, the determination of the intrinsic correlations between the luminosities is not straightforward.  It is affected by the observational selection effects and other factors that truncate the data, sometimes in a complex manner [\cite[Antonucci (2011)]{Ski} and \cite[Pavildou et al. (2010)]{Pavildou12}].  In this paper we describe  methods that allow us to determine the evolution of the radio and optical luminosities, and determine the true intrinsic correlation between the two luminosities.  We find a much weaker correlation than  observed  and sub-linear relations between the luminosities. This has a significant implication for jet and accretion disk physics.

\firstsection 
\section{Introduction}\label{intro}

It is generally believed that the radio and gamma-ray emission of active galactic nuclei (AGNs) come from their jets while the optical (UV and possibly X-ray) emission comes from the accretion disk. Thus, investigation of the relative emissions in the optical and radio bands can shed light on the interrelated processes forming the jet and the disk. There are two distinct and complementary methods of treating this problem. The first and most common method is to carry out  detailed modeling of the emission characteristics over a wide band of frequencies for individual sources. By its nature, this method is limited to only bright and relatively nearby sources, with simultaneous observations at all wavelengths, so that the information it provides may be biased and not representative of the population as a whole. A second method is to investigate the statistical properties of the population as a whole and determine, among other things, the correlation between the characteristics of the two emissions. The second approach will be the focus of this paper. We often see plots of one emission characteristic vs another.  The most elementary plots are those involving the measured fluxes (e.g optical and radio $f_o, f_r$), which often show some correlation. Such correlations are difficult to interpret because a flux is not an intrinsic characteristic.  For sources with measured  redshifts ($z$) one can calculate the monochromatic luminosities 
\ as
\beq\label{L}
L_{o,r}=4\pi d^2_L(z)(z)f_{o,r}/K_{o,r}, 
\eeq
where $d_L$ is luminosity distance and the K-correction $K(z)\sim (1+z)^{1+\alpha}$, for a power law spectral index of $\alpha$.
Plots of luminosities vs. each other obtained this way are also very common and often show large correlation coefficients  $r_{L_oL_r}$). As is known widely (but apparently not universally) some of this correlation is induced by the strong dependence of both quantities on the distance.\footnote{This is not limited to $L-L$ correlation nor to extragalactic or high redshift sources.  It applies to any two variable which depend on a third one (here the distance) with a large fractional range.}

Our focus here will be the jet-disk connection and the $L_o-L_r$ correlation. In \S \ref{over} we give an overview of the problem and in \S \ref{method}  a brief description of the methods we use. The results from application of these methods are discussed in \S \ref{result}, where we use a sample of $> 5000$ quasars from the joint Sloan Digital Sky Survey (SDSS) [\cite[Schneider et al. (2010)]{SDSSQ}; 65,000 sources with i mag. $<$19.1] and Faint Images of the Radio Sky at Twenty centimeters (FIRST) [\cite[Becker et al. (1995)]{FIRST1}); 300,000 sources with 1.4 GHz flux $>1$ mJy] data, compiled in \cite[Singal et al. (2013)]{QP2}. In \S \ref{sum} we give a summary and conclusions.

\begin{figure}[b]
\begin{center}
\includegraphics[width=6.5cm]{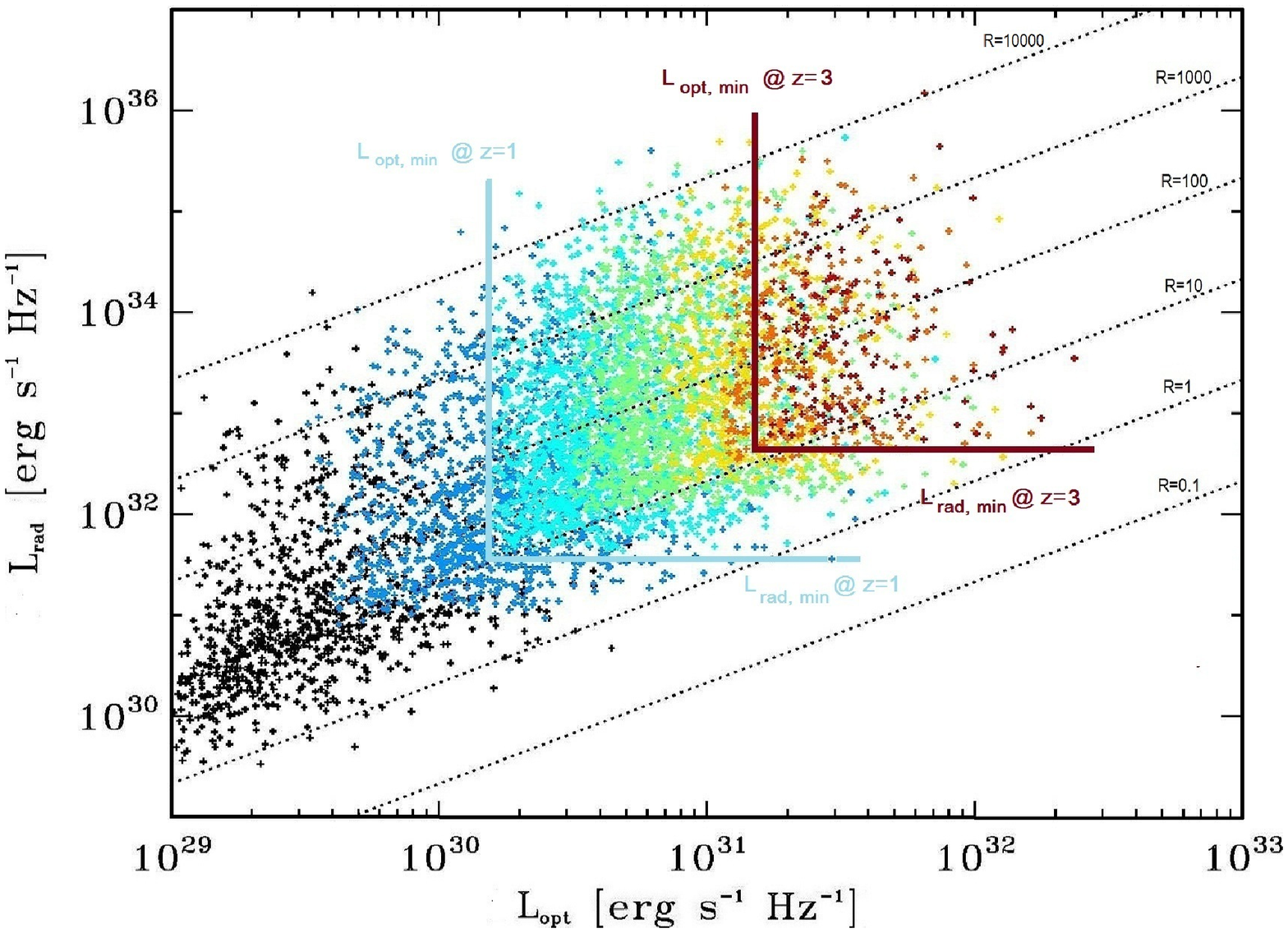}
\includegraphics[height=5.0cm,width=6.5cm]{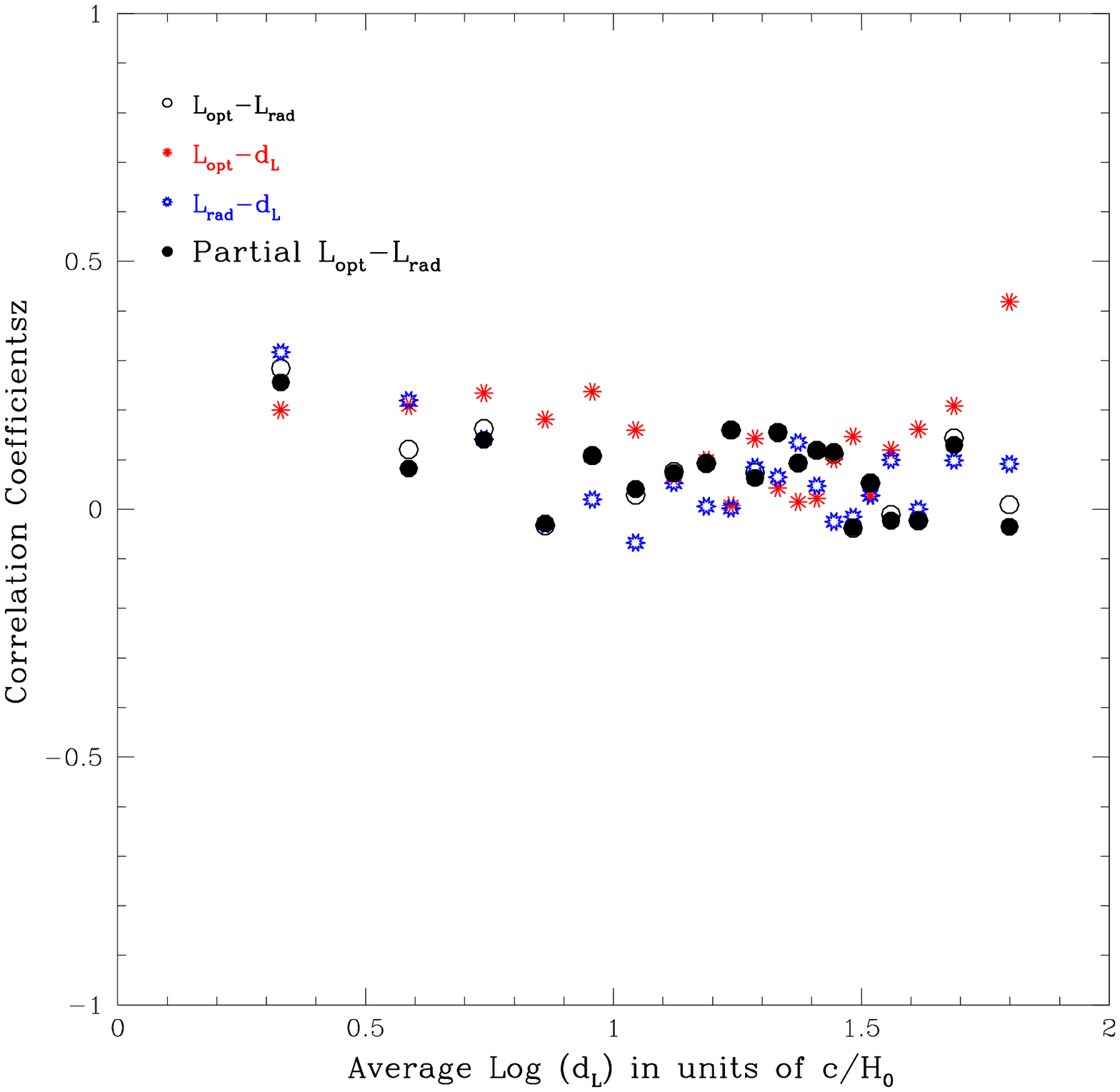}
 \caption{{\bf Left:} Radio (1.4 GHz) vs optical (2500 \AA) luminosity of the SDSSxFIRST sample of quasars taken from \cite[Singal et al. (2013)]{QP2}, with binned redshifts. The total sample has a correlation coefficient of 0.68 but as shown in the {\bf right panel} the ordinary (open circle) and partial (filled circle) correlation coefficients in bins with small ranges of distances are insignificant. The red and blue points show the binned luminosity-distance correlation coefficients.}
   \label{LLCorr}
\end{center}
\end{figure}

\section{Overview}\label{over}

\subsection{Effects of the Distance Dependence}

Fig.\,\ref{LLCorr} (left) shows a plot of radio vs optical luminosities from the SDSSxFIRST sample showing a strong correlation  $L_r\propto L_o^{1.2}$  and a correlation coefficient of $r_{L_oL_r}=0.68$, which is  highly significant.  However, the color coding in this figure shows that the correlation is induced by the distance dependence. The right panel shows that the correlation coefficients for individual equal number bins are reduced to insignificant levels, indicating that the raw correlation of the whole sample is not a measure of the true correlation.

In terms of logarithm of the quantities  Eq.\,\ref{L} simplifies to a linear relation
\beq\label{defs}
X_i= 2\times Z+B_i\,\,\,\,{\rm where} \,\, X_o=\ln L_o,\,\, X_r=\ln
L_r\,\,{\rm and}\,\,Z=\ln d_L,
\eeq  in which case  one can then use the ``partial correlation coefficient''
\beq\label{partial}
r_{X_oX_r;Z}={r_{X_oX_r}-r_{X_oZ}r_{X_rZ}\over  (1-r^2_{X_oZ})(1-r^2_{X_rZ})},
\eeq
which accounts for mutual distance dependence of the luminosities.  The partial correlation coefficients for the sample is smaller
$r_{L_oL_r;d_L}=0.21$, which  considering the large number of data points is still significant. As expected, for the binned data shown in Fig.\,\ref{LLCorr} (right), the partial and ordinary correlations have similar values.

\begin{figure}[b]
\begin{center}
\includegraphics[width=6.5cm]{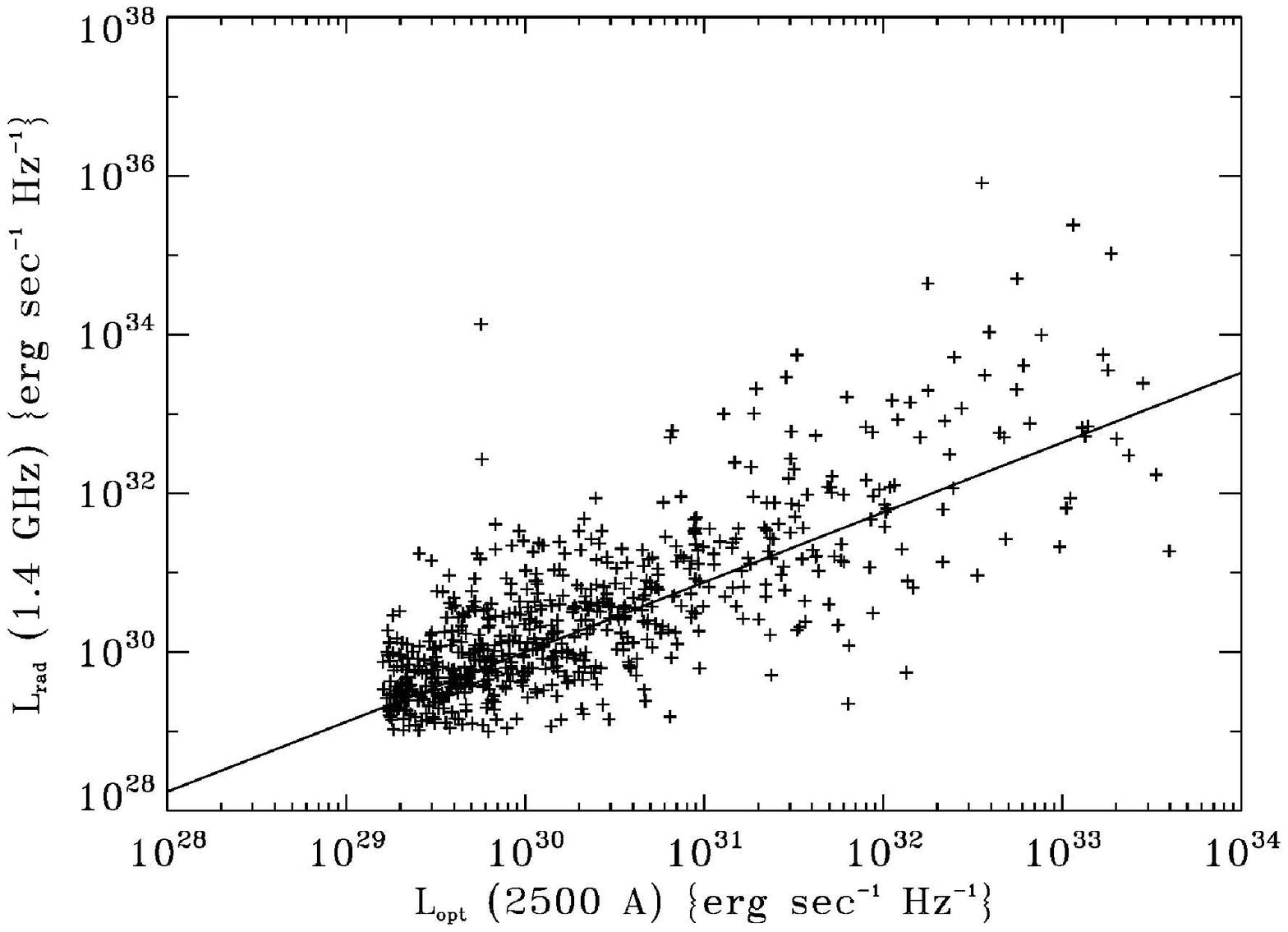}
\includegraphics[width=6.5cm]{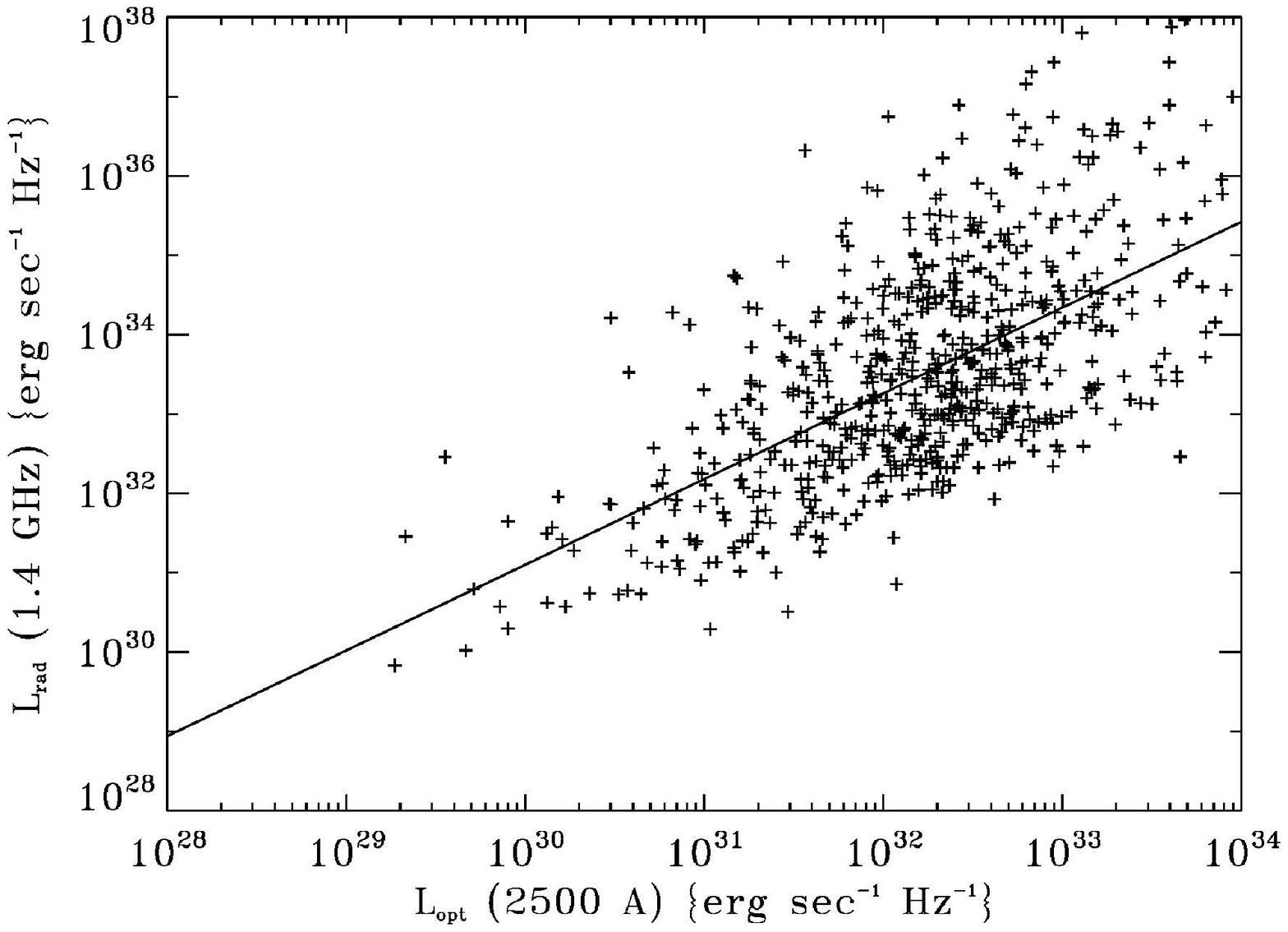}
 \caption{Radio vs optical luminosity scatter diagrams for two flux limited simulated observational samples of quasars with characteristics similar to that of the observed quasars but without any luminosity evolution. The lines show power law
fits $L_r\propto L_o^\alpha$. {\bf Left:} Uncorrelated luminosities showing correlation with $\alpha=0.88\pm 0.03$. {\bf Right:}
Linearly correlated luminosities showing  $\alpha=1.08\pm 0.05$.}
   \label{sim}
\end{center}
\end{figure}

To further demonstrate the importance of some of these effects (and provide an approximate measure) we have simulated two samples of sources with luminosity correlation $L_r\propto L_o^\alpha$; in the first the two luminosities were uncorrelated ($\alpha=0$) and in the second they have a linear correlation ($\alpha=1$). The other characteristics like the shapes of the luminosity function (LF) and  density evolution were chosen to mimic those of the observed quasars, except we did not include any luminosity evolution.  First we generated a large number of sources and then selected a radio and optical flux limited sample. The two panels of Fig.\,\ref{sim} show the $L-L$ scatter diagrams for the two resultant samples. It is clear that the correlated sample shows a stronger correlation, with $\alpha \sim 1.1$, that is significantly larger than one, and that the uncorrelated sample also shows some significant correlation $\alpha \sim 0.9$.  A cursory comparison of these scatter diagrams with that of the  observed sample (Fig.\,\ref{LLCorr}, left), which has an $\alpha \sim 1.2$ would imply that the radio and optical luminosities of quasars are approximately linearly correlated. As described below this is not the case because, unlike the simulated quasars, the observed quasars show a substantial luminosity evolution. 

\subsection{Effects of Luminosity Evolution}

\begin{figure}[b]
\begin{center}
\includegraphics[width=6.5cm]{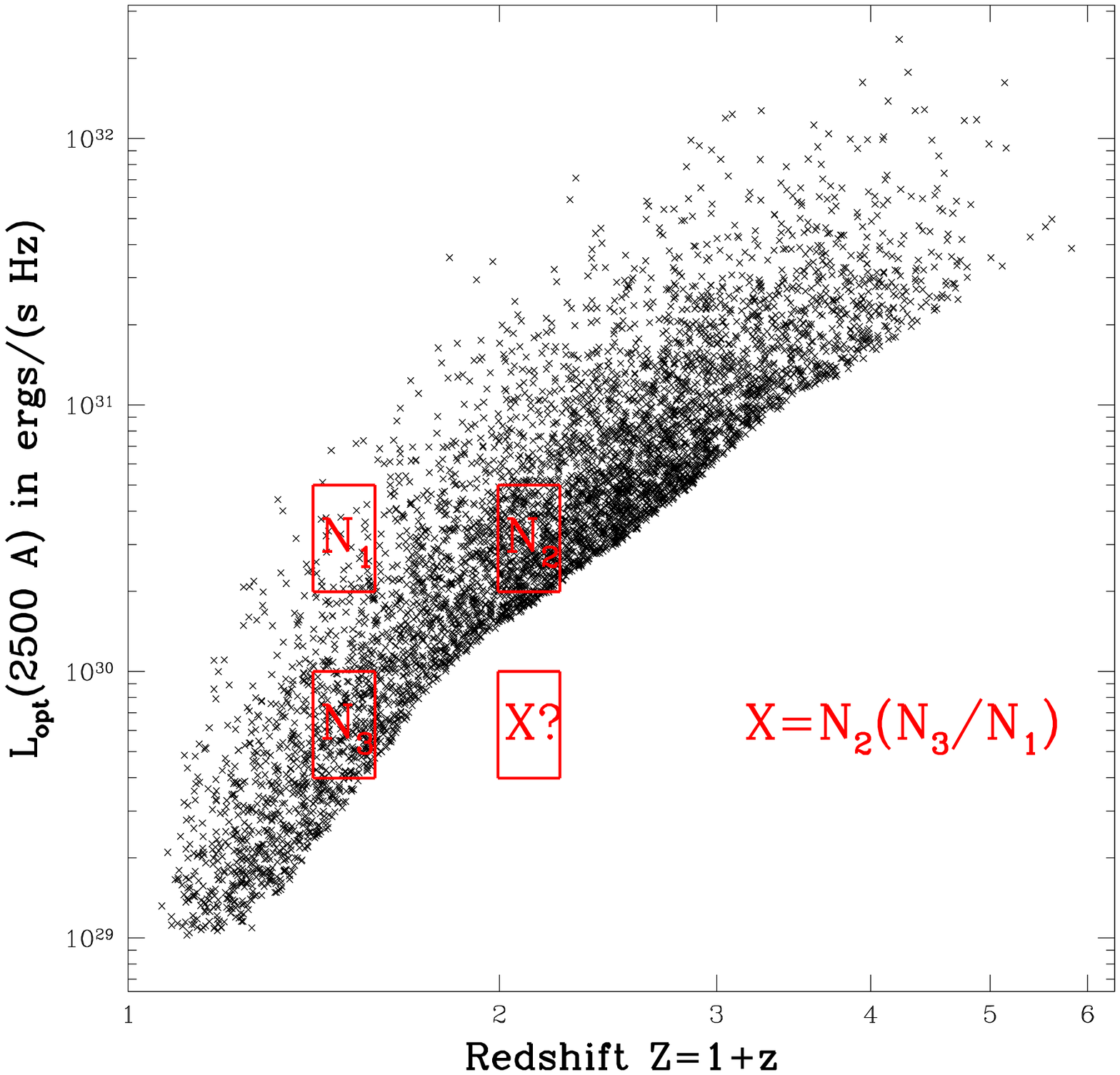}
\includegraphics[width=6.5cm]{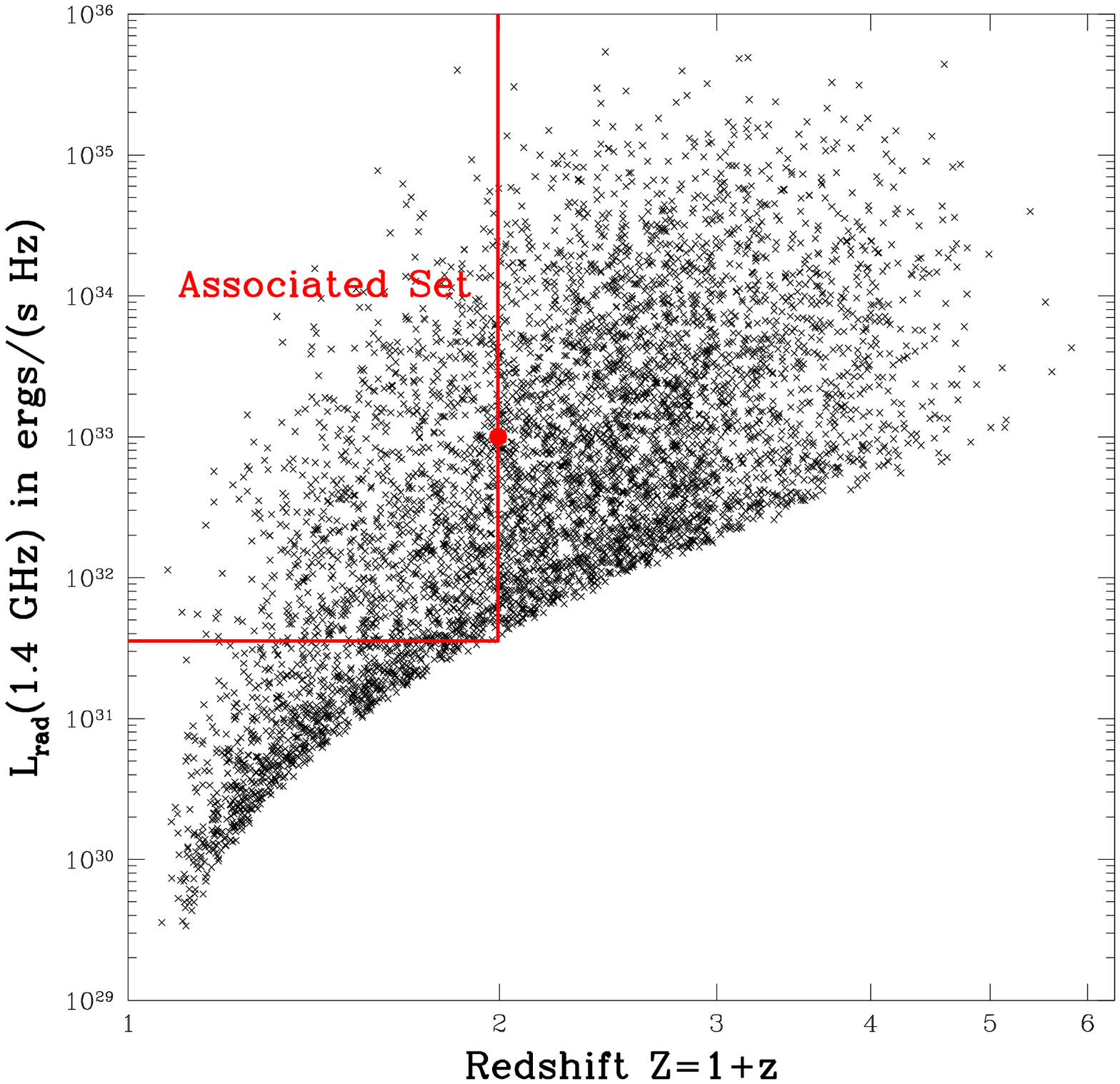}
 \caption{Optical luminosity at 2500 \AA ({\bf Left}) and radio luminosity at 1.4 GHz ({\bf Right}) versus redshift of the SDSSxFIRST sample. These data are truncated from below; $L>4\pi d^2_L(z)f_{\rm min}/K(z)$. The red lines visualize the {\it associated set} of the data point (red dot). The red squares demonstrate that in case of independence the numbers of sources in any box in the truncated region can be obtained from those in equivalent boxes with observed sources.}
   \label{LdCorr}
\end{center}
\end{figure}

Thus partial correlation analysis does not  solve the problem completely because it requires a knowledge of the true correlations, $r_{X_oZ}, r_{X_rZ}$, of the luminosities with distance, $r_{X_oZ}, r_{X_rZ}$, which ordinarily is called the luminosity evolution.  In general, the sample of sources used for population studies are subject to various observational selection effects which truncates the data. The simplest truncated samples, such as the ones we will consider here, are flux limited samples ($f_{o,r}>f_{\rm min_{o,r}}$).  Those shown in Fig.\,\ref{LdCorr} clearly show very strong {\it observed} $L-z$ correlations (roughly $L_o\propto (1+z)^{\sim 5}; L_r\propto (1+z)^{\sim 8}$) with correlation coefficients $r_{L_od_L}\sim 0.9$ and $r_{L_od_L}\sim 0.7$.  Most of this correlation is due to the flux limit.  However, as is well known, AGNs undergo a substantial luminosity evolution and, more importantly, the luminosity evolutions in radio and optical bands are different. The red and blue points in the right panel of Fig.\,\ref{LLCorr} show the binned values of $r_{L_od_L}$ and $r_{L_od_L}$ which are smaller than those of the total sample but still significant.

The simple correlations  described above would provide a true measure of the  $L-L$ correlation if there were no data truncation and no luminosity evolutions. To obtain the true measure one must account for both effects.  {\it The General solution} of this  problem requires a simultaneous determination of the unbiased intrinsic correlations between all three variables with the proper accounting of the data truncations due to observational selection effects; i.e. the  problem is to provide a complete description of the \underline{tri-variate distribution} $\Psi(L_o,L_r,z)$;  the joint radio and optical LF and its evolution.

\section{Methods and Approach}\label{method}

We will demonstrate  our approach for determination of single LF and its evolution;  $\Psi(L,z)$ from a flux limited sample. Fig.\,\ref{LdCorr}  shows  two such flux limited samples with $L\geq 4\pi d_L^2f_{min}/K$. The bias introduced here is known as the {\bf Malmquist Bias}. Many papers have been written to correct for this bias in astronomical literature [\cite[Malmquist (1925)]{Malmquist25}; \cite[Eddington (1940)]{Eddington40};  \cite[Trumpler \& Weaver (1953)]{TW53}].  Most of these early procedures assumed simple parametric forms (e.g Gaussian) distributions.  However, since the discovery of quasars  the tendency has been to use non-parametric methods: e.g $\langle V/V_{\rm max}\rangle$ [\cite[Petrosian (1973)]{P73} and \cite[Schmidt (1972)]{Schmidt72}] or the $C^{-}$ (\cite[Lynden-Bell 1971]{L-B71}). For a more detailed description see review by \cite[Petrosian (1992)]{P92}. All these methods however, suffer from the major shortcoming that they assume that the variables are \underline{independent or uncorrelated}; $\Psi(L,z)=\psi(L)\rho(z)$, i.e. there is no luminosity evolution; an assumption that is often incorrect, especially for AGNs. Thus, the first task should be the determination of the correlation between the variables. A commonly used non-parametric method is  the Spearman Rank test. This method, however, fails for a truncated data set. Efron \& Petrosian developed a novel
method to account for this truncation; \cite[Efron \& Petrosian (1992)]{EP92} and \cite[Efron \& Petrosian (1999)]{EP99}.  The gist of the method is to determine the rank  $R_i$ of each data point among its {\it associated  set}, namely the largest un-truncated subset (e.g. one shown by the red lines in the
right panel of Fig.\,\ref{LdCorr}). Then using a test statistic, e.g. Kendell's tau defined as $\tau=\Sigma_i(R_j-E_j)/\sqrt{\Sigma_i V_j}$ one can determine
the degree of correlation. Here $E_j$ and $V_j$ are the expectation and variance of the ranks.  A small value ($\tau\ll 1$) would imply independence, and
$\tau>1$ would imply significant correlation. One can then redefine the variables, e.g. define a ``local" luminosity $L'=L/g(z)$ [with $g(0)=1$] using a parametric form, e.g $g(z)=(1+z)^k$, and calculate $\tau$ as a function of the parameter $k$. The values of $k$ for which $\tau=0$ and  $\tau=\pm 1$ give the best value and one sigma range for independence.  

Once independence is established then one can use above mentioned methods to determine the mono-variate distributions; local LF $\psi(L')$ and the density
evolution $\rho(z)$. Fig.\,\ref{LdCorr} (left) demonstrates this step. If the variables are independent, then one can build up a complete description of the
distribution from the number $X$ of truncated sources in the specified box. We, however, use no such binning but build up both distributions point by point based on the $C^{-}$ methods and the concept of the {\it associated set}.  These procedures have been used for analysis of cosmological evolutions of quasars in \cite[Maloney \& Petrosian (1999)]{MP99}, \cite[Singal et al. (2011)]{QP1} and \cite[Singal et al. (2013)]{QP2}, gamma-ray bursts in \cite[Lloyd et al. (2000)]{Lloyd00}, \cite[Kocevski \& Liang (2006)]{KL06}, and \cite[Dainotti et al. (2013)]{MD}, and blazars in \cite[Singal et al. (2012)]{BP1} and \cite[Singal et al. (2014)]{BP2}. 

\section{Luminosity-Luminosity-Redshift Correlations}\label{result}

We now show results from applications of the above methods to the SDSSxFIRST sample introduced above. The complication in this case is that we are dealing with a tri-variate distribution $\Psi(L_o,L_r,z)$ and need to determine the degree  of the three correlations. In what follows we will use simple one or two parameter correlation forms $L_o=L'_o \times g_o(z)=L_o(1+z)^{k_o}/h_o(z), L_r=L'_r \times g_r(z)=L'_r(1+z)^{k_r}/h_r(z)$ and $L^{cr}_r=L_r(L_0/L_o)^\alpha$.  Here super script $^{cr}$ stands for ``correlation reduced'', $L_0$ is some fiducial luminosity and as in the past we adapt $h_i(z)=1+[(1+z)/Z_{i,cr}]^{k_i}$ with $Z_{cr,i}$ a fiducial redshift ($>1$) which allows a high redshift cutoff of the luminosity evolutions. Ideally one would like a simultaneous fitting to all these parameters finding their best values and one sigma ranges. In the {\it Singal et al.} papers papers we used two approaches. In first we assumed that all of the observed  $L-L$ correlation is intrinsic and determined the value of $\alpha \sim 1.2$ mentioned above and then proceeded to find the luminosity evolutions of $L_o$ and $L^{cr}_r$ which were then translated to  evolution of radio and optical luminosity evolution parameters ($k_o, k_r$). In the second  we assumed that all of the observed  $L-L$ correlation is  induced by observations and determined luminosity evolution parameters directly. The luminosity evolutions were determined simultaneously by finding the minimum of the combined Kendell tau $\tau=\sqrt {\tau_o^2 + \tau_r^2}$, using a constant value for $Z_{cr,i}=3.7$. We obtained similar results with both methods. 

\begin{figure}[b]
\begin{center}
\includegraphics[width=6.5cm]{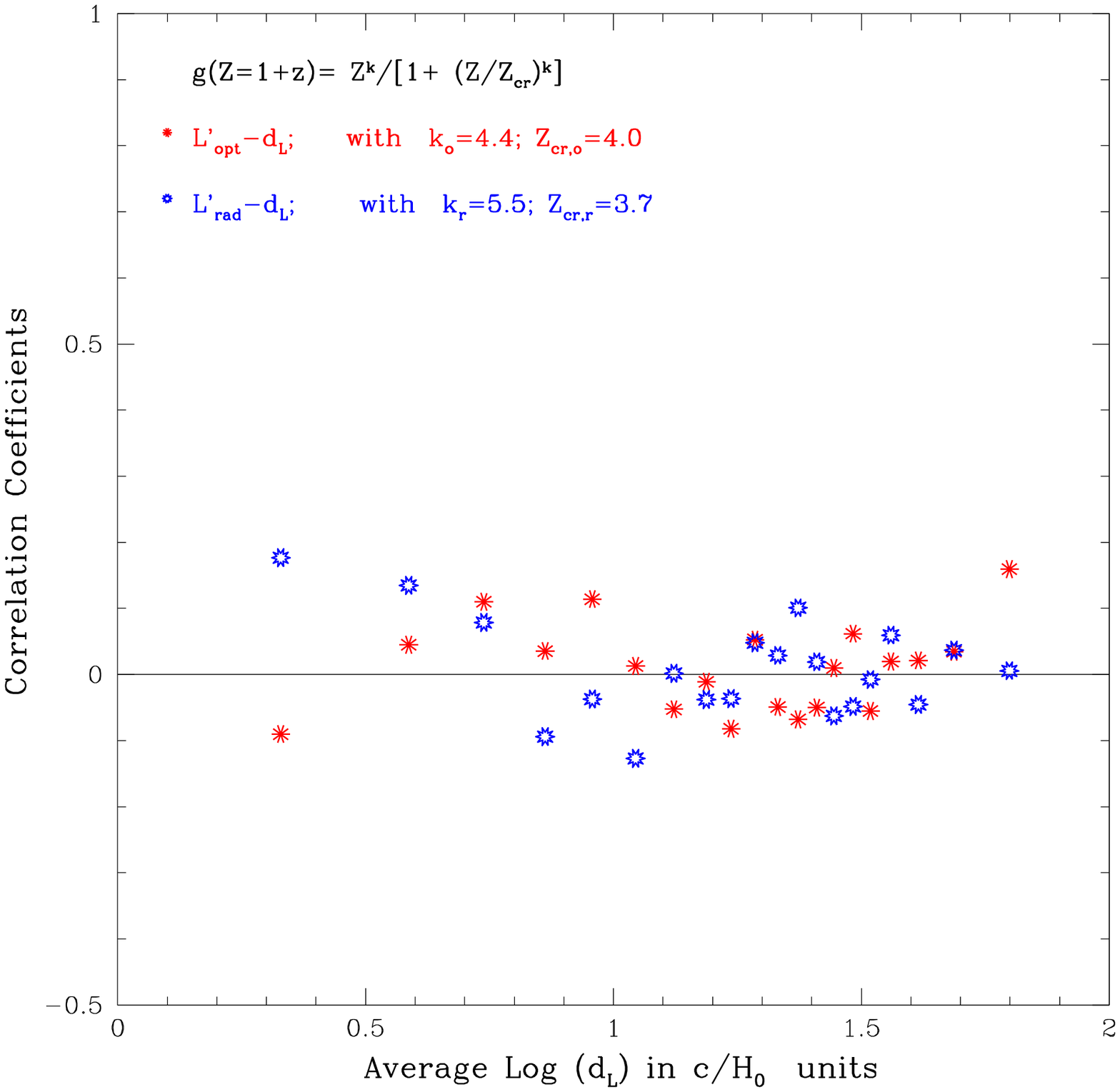}
\includegraphics[width=6.5cm]{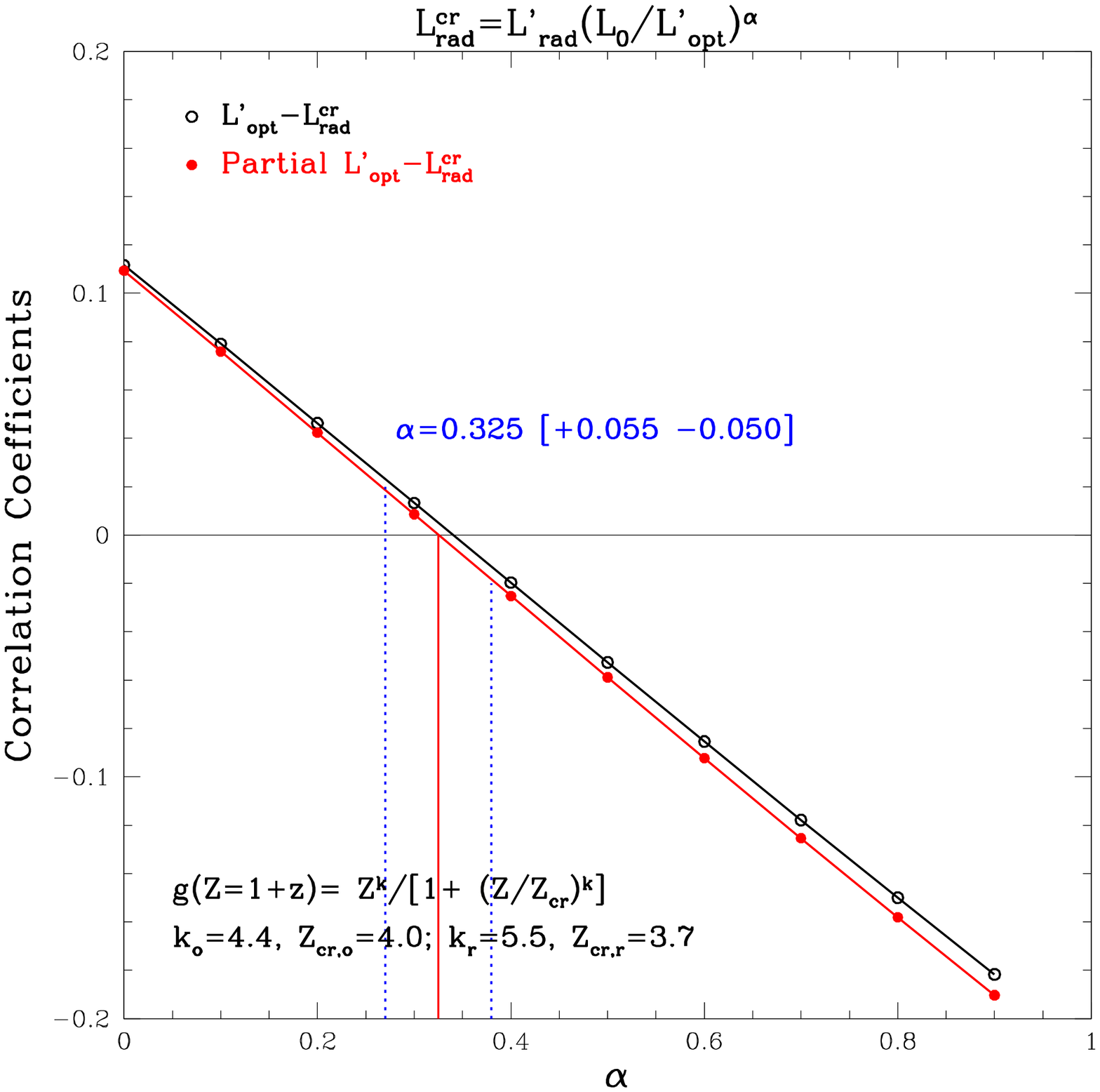}
 \caption{{\bf Left:} $L-d_L$ correlation  coefficients $r_{L_od_L}$ (filled red) and $r_{L_rd_L}$ (open blue) versus average luminosity distance, for the
indicated evolution parameters, showing relatively insignificant correlation at almost all redshifts.  The values of the parameters $k_i$, and $Z_{cr,i}$ which remove the correlation between luminosity and distance are the ones which best describe the luminosity evolution.  {\bf Right:}  Local luminosity-luminosity correlation coefficient vs  correlation index $\alpha$ showing the best value and the range outside of which the probability that $L^{cr}_r$ and $L'_o$ are correlated is significant.}
   \label{results}
\end{center}
\end{figure}

In what follows we start with the above derived luminosity evolution parameters and using an iterative procedure try to get a consistent picture for all three correlations. The one-parameter forms provide a satisfactory measure of the global evolution (averaged over all redshifts). As evident from Fig.\,\ref{LLCorr} (right), the binned raw  coefficients $r_{L_od_L}$ and $r_{L_rd_L}$ are significant and vary considerable with  distance.  We have used a two parameter evolutionary form, varying both $k_i$ and $Z_{cr,i}$, and determined a narrow range of parameter values for which the total correlation coefficients are small.  Not only that but as shown in Fig.\,\ref{results} (left) for the specified set of parameters the binned values are also sufficiently small (so that the probability that each binned set is drawn from a random sample is large) and show only minor variations. Obviously one can do better by the use of  evolution forms with three or more parameters, but this is beyond the scope here. 

Now using these evolutions we  transfer the observed luminosities $L_o, L_r$ to their local values $L'_o,L'_r$, whose distribution  show much  weaker correlation. However, the correlation coefficient though small is still significant. To quantify this  we now return to the \underline{local} correlation reduced radio luminosity $L^{cr}_r=l'_r(L_0/L'_o)^\alpha$ and calculate the $r_{L^{cr}_rL_o}$ correlation coefficient as a function of index $\alpha$. As shown in  the right panel of Fig.\,\ref{results}, the correlation becomes insignificant for $\alpha=0.325\pm 0.05$. Thus, we have parameterize all correlations using some simple but useful relations.

\section{Summary and Conclusions}\label{sum}

We have pointed out that one can approach the relation between jet and accretion disk emissions of AGNs using investigation of the correlations between jet and disk luminosities,  as represented here by radio and optical observations. We point out that the raw observed correlations are far from the true intrinsic values
because they are affected by the mutual dependence of the luminosities on the distance and because of considerable (and different) luminosity evolution of AGNs in different bands.  We have described methods that can account for this effects given that the available flux limited data are strongly truncated. 

We first determine the radio and optical luminosity evolutions using two parameters for each band. This allows us to obtain a local luminosity-luminosity distribution. We use a simple power law form for the luminosity evolutions and find a best value for the power law index $\alpha=0.32$ which is much smaller than the value of $\sim$1.2 obtained from the raw data. These results indicate that locally the luminosities are correlated but not linearly as one may naively expect because both jet and disk emissions are governed by the black hole mass $M_{BH}$ and the accretion rate ${\dot M}$. This may indicate other factors, such as the spin of the black hole  and/or the associated magnetic fields also play an important role. 

We have determined the correlation index for the local luminosities. But, because the two luminosities evolve differently we expect the $L-L$ correlation also to evolve. How the fiducial luminosity $L_0$ and the exponent $\alpha$ evolve requires further investigation.  A more robust way of carrying out this work would be to start with the local luminosities including the evolution function  $g_i(z)$, which would make the term $B_i$ in Eq. (\ref{defs}) also dependent on $g_i(z)$, and define a general correlation reduced luminosity $L^{cr}=G(L_r, L_o)$. Then one should simultaneously minimize some statistics (like Kendell's tau or correlation coefficient) to obtain the best values of all parameters of the correlation functions $g_i(z)$ and $G(L_r, L_o)$. This will be explored in future works.

\end{document}